\documentclass[12pt,preprint]{emulateapj}
\usepackage{epsfig}
\usepackage{amsmath}
\usepackage{natbib}

\def \teff {$T_{\rm eff}$}
\def \logteff {${\rm log}(T_{\rm eff})$}
\def \logg {$\log (g)$}
\def \ttau {$T$-$\tau$}

\def \rev {}

\begin{document}


\submitted{}

\title{The Effect of Metallicity-Dependent {\ttau} Relations on Calibrated Stellar Models}
\author{Joel D. Tanner, Sarbani Basu \& Pierre Demarque}
\affil{Department of Astronomy, Yale University, P.O. BOX 208101, New Haven, CT 06520-8101}
{\email{joel.tanner@yale.edu}}


\begin{abstract}

Mixing length theory is the predominant treatment of convection in stellar models today.  Usually described by a single free parameter, $\alpha$, the common practice is to calibrate it using the properties of the Sun, and apply it to all other stellar models as well.  Asteroseismic data from \textit{Kepler} and \textit{CoRoT} provide precise properties of other stars which can be used to determine $\alpha$ as well, and a recent study of stars in the \textit{Kepler} field of view found $\alpha$ to vary with metallicity.  \rev{Interpreting $\alpha$ obtained from calibrated stellar models}, however, is complicated by the fact that the value for $\alpha$ depends on the surface boundary condition of the stellar model, or {\ttau} relation.  Calibrated models that use typical {\ttau} relations, which are static and insensitive to chemical composition, do not include the complete effect of metallicity on $\alpha$.  We use 3D radiation-hydrodynamic simulations to extract metallicity-dependent {\ttau} relations and use them in calibrated stellar models.  We find the \rev{previously reported} $\alpha$-metallicity trend to be robust, and not significantly affected by the surface boundary condition of the stellar models.

\end{abstract}

\section{Introduction} \label{sec:introduction}

The Mixing Length Theory \citep[MLT;][]{1958ZA.....46..108B} remains one of the most popular treatments for stellar convection, and typically describes convective eddy sizes with a single parameter, $\alpha$, which is arbitrarily set by the modeler.  Generally, $\alpha$ is kept fixed at a value that is determined by calibrating \rev{a solar model} to the properties of the Sun, however, this  has no \textit{a priori} justification.  With space-based photometric data from \textit{CoRoT} \citep{corot} and \textit{Kepler} \citep{kepler}, it is now possible to precisely constrain the properties of other stars, \rev{which allows $\alpha$ to be determined in a similar manner.  Determining}
$\alpha$ is possible with tight constraints on the stellar mass and radius, however, it still  depends on a variety of input microphysics and boundary conditions.

It is increasingly apparent that the usual approach of using a constant solar-calibrated value for $\alpha$ is not appropriate for stars that differ from the Sun, either in composition, mass, or stage of evolution.  As an example, asteroseismic studies of $\alpha$ Cen \citep[e.g.][]{dem86,fer95,mig05} require a non-solar value of the mixing length parameter to reproduce the stellar radius.  Stars from \textit{CoRoT} and \textit{Kepler} also require non-solar mixing length values to model the oscillation spectra \citep[e.g.,][]{met10, deh11, mat12}.  

More recently, \citet{2012ApJ...755L..12B} found a systematic metallicity dependence of $\alpha$ with calibrated stellar models of stars in the Kepler field of view.  It is challenging to interpret the meaning of $\alpha$ in calibrated stellar models, since it is intertwined with the atmospheric boundary condition and other input microphysics.   The purpose of this study is to test whether the trends reported by \citeauthor{2012ApJ...755L..12B} are sensitive to the treatment of the surface boundary condition, which is expected to vary with metallicity. \rev{Note that in the present work, we do not provide a calibration of the mixing length parameter, but rather we investigate how a metallicity-dependent surface boundary condition would affect such a calibration.}  

Adjusting the value of $\alpha$ determines the specific entropy of the convection zone, which in turn sets the stellar radius.  In effect, $\alpha$ is a free parameter that defines the radius of the stellar model.  The effect of the MLT on the stellar model is most important near the surface, where the specific entropy of the convection zone is set.  This so-called superadiabatic layer (SAL) is several scale heights above and below the stellar photosphere, defined in stellar models as where $T=T_{\rm eff}$.  Stellar models usually separate the atmosphere (defined as the layers above the {\teff} surface) from the stellar envelope, and provide an atmospheric structure from a {\ttau} relation \rev{which is integrated inward from a very small optical depth}.    This relation, along with the assumption of hydrostatic equilibrium, defines the structure of the outermost layers of the stellar model.

In the \citeauthor{2012ApJ...755L..12B} study, all of the models were computed with a fixed {\ttau} relation, which is the usual approach.  However, since the surface boundary condition is determined in part by the nature of the near-surface stellar convection, the relationship between the value of the mixing length parameter and metallicity may be affected by the arbitrary choice of {\ttau} relation.  In the following sections we describe several tests conducted using stellar models computed with a variety of surface boundary conditions;  these range from the usual prescribed static {\ttau} relations to using atmospheric stratifications extracted from 3D simulations.

\section{{\ttau} Relations in This Work} \label{sec:ttau}

The typical approach to treating the surface boundary in stellar models is to impose an atmospheric structure, or {\ttau} relation.  One of the more popular boundary condition is the Eddington {\ttau} relation, which is purely radiative and does not include any effect from photospheric convection or overshoot.  In the Eddington {\ttau} relation, the boundary of the photosphere is fixed at an optical depth of $\tau=2/3$, and the relation is not sensitive to variation in metallicity.  Other popular alternatives to the Eddington {\ttau} include the semi-empirical KS \citep{1966ApJ...145..174K} and VAL relations \citep{1981ApJS...45..635V}.  The choice of {\ttau} relation directly affects the value of $\alpha$ when constructing a model with a particular mass and radius.  For example, standard solar models computed with the Eddington and KS {\ttau} relations have mixing length parameters of approximately $1.8$ and $2.1$, respectively.

Semi-empirical {\ttau} relations, such as the KS \citep{1966ApJ...145..174K} and VAL \citep{1981ApJS...45..635V} relation, can be used instead of the purely radiative Eddington relation. These relations are derived from the Sun, and so are applicable to models with the solar composition, mass, and radius.  While they are likely to be an improvement for computing solar models, they are, however, not necessarily any better for models of stars other than the Sun.  

One way to get metallicity-dependent {\ttau} relation is through 3D radiation hydrodynamic simulations. Simulations provide a realistic and self-consistent description of stellar convection by following the gas dynamics from the near-adiabatic region below the SAL to the optically thin atmosphere.  In a simulation there is no distinction between the atmosphere and interior as there is in a stellar model, and the effect of convective overshoot is naturally included in the simulated stratification. Simulations of photospheric convection show a range of convective properties across the HR diagram \citep[e.g.][]{2011ApJ...731...78T, 2013arXiv1302.2621M} and with chemical composition \citep{2013ApJ...767...78T, 2013ApJ...778..117T}.  Simulations also reveal that in addition to changing the convective properties, metallicity also changes the atmospheric stratification in ways that are not represented by the static {\ttau} relations used in 1D stellar models \citep[e.g.][]{1999A&A...346L..17A, 2013ApJ...767...78T}.

We extract {\ttau} relations from a grid of simulations at a fixed surface gravity ($\log (g)=4.30$), but span a range in effective temperature, and are divided into four groups according to metallicity.   The basic properties of the simulations are summarized in Table \ref{tab:sims}, and further details can be found in \citet{2012ApJ...759..120T}. The range in effective temperature is comparable to the stars in the \citet{2012ApJ...755L..12B} data set, and the metallicity variation extends to lower-Z.  \rev{The grid comprises simulations with four metallicities, ranging from slightly super-solar ($Z=0.040$) to very sub-solar ($Z=0.001$).  The precise Solar composition is not included, although the $Z=0.020$ simulations  have roughly the same metallicity as the Sun.} Note that the mass-radius relation for the corresponding calibrated models is fixed because of the {\logg} constraint.  The helium mass fraction (Y) is held constant, and the hydrogen mass fraction (X) adjusted according to the change in metallicity (Z).

The simulation domain is a Cartesian box, and the temperature stratification ($T$) from each simulation is extracted by taking temporal and spatial averages as a function of optical depth ($\tau$).  For a given temporal snapshot, the optical depth is calculated for each vertical column by integrating the opacity and density:
\begin{equation}
\tau = \int \rho(z) \kappa(z) dz
\end{equation}
This results in many $T(\tau)$ profiles (one for each column) which are spatially averaged by interpolating onto a uniform $\tau$ grid.  This is repeated for many snapshots spread uniformly over several thermal timescales, until statistical convergence is achieved.

\begin{table}
  \caption{Global Properties of 3D Simulations}
  \label{tab:sims}
  \begin{center}
    \leavevmode
    \begin{tabular}{lllll} \hline \hline              
  $ID$          & Z              & X & $\log T_{\rm eff}$      \\ \hline 
s1 & 	0.040 & 	0.715 & 	3.694  \\
s2 & 	0.040 & 	0.715 & 	3.716  \\
s3 & 	0.040 & 	0.715 & 	3.737  \\
s4 & 	0.040 & 	0.715 & 	3.757  \\
s5 & 	0.020 & 	0.735 & 	3.709  \\
s6 & 	0.020 & 	0.735 & 	3.730  \\
s7 & 	0.020 & 	0.735 & 	3.750  \\
s8 & 	0.020 & 	0.735 & 	3.770  \\
s9 & 	0.010 & 	0.745 & 	3.726  \\
s10 & 	0.010 & 	0.745 & 	3.746  \\
s11 & 	0.010 & 	0.745 & 	3.764  \\
s12 & 	0.010 & 	0.745 & 	3.780  \\
s13 & 	0.001 & 	0.754 & 	3.759  \\
s14 & 	0.001 & 	0.754 & 	3.771  \\
s15 & 	0.001 & 	0.754 & 	3.783  \\
s16 & 	0.001 & 	0.754 & 	3.795  \\ \hline
    \end{tabular}
  \end{center}
    Properties of 3D RHD simulations from \citet{2012ApJ...759..120T}.  All simulations have the same surface gravity ($\log g = 4.30$).  There are four metallicity groups with overlapping ranges in {\teff}.
\end{table}

\begin{figure}[h]
\epsscale{1.15}
\plotone{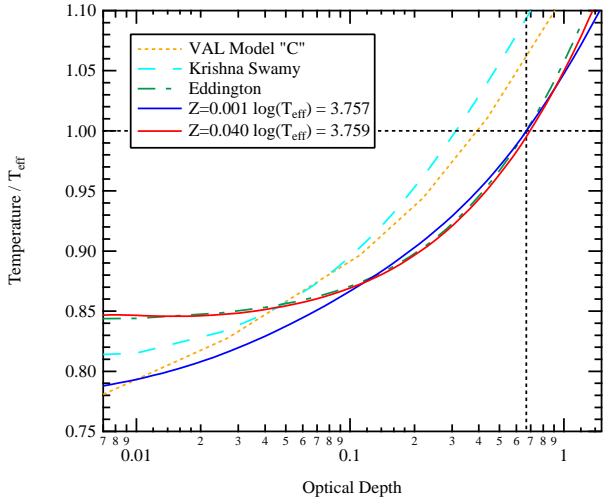}
\caption{The Eddington {\ttau} relation compared with semi-empirical {\ttau} relations for the Sun.  Also shown are average {\ttau} relations taken from 3D RHD  for simulations with varied metallicity and fixed {\logg} and {\logteff}.}
\label{fig:ttau}
\end{figure}

Each of the simulated {\ttau} relations uniquely represent the thermal structure of a star with the corresponding surface gravity ({\logg}=4.30), effective temperature ({\logteff}) and composition.  Figure \ref{fig:ttau} compares the {\ttau} relations that were used to compute the calibrated stellar models.  They differ both in the nearly isothermal region at low optical depth, and near the photosphere which is defined as where $T=T_{\rm eff}$.  In particular, the photosphere of the semi-empirical {\ttau} relations is shifted to smaller optical depth relative to the the Eddington relation.  The value of the mixing length parameter is particularly sensitive to the location of the photospere because the stellar structure equations are integrated inward from this point.  

Radiative heating and cooling plays an important role in the optically thin layers.  An important source of cooling, which is neglected in the Eddington {\ttau} relation, is the adiabatic cooling from the rising and expanding convective granules, i.e., atmospheric overshoot.  The temperature structure in the optically thin layers is the result of balancing radiative heating with adiabatic cooling.  As the simulations can more accurately account for the additional cooling, their {\ttau} relations tend to approach a cooler isotherm.  The temperature of the optically thin layers also depends on the metallicity, with lower opacities resulting in steeper {\ttau} gradients and cooler atmospheres.  The Eddington, KS, and VAL relations are insensitive to chemical composition, and so are identical in models with varied metallicity.  We use our grid of 3D simulations to provide simulated {\ttau} relations that are used for modeling stars with the corresponding {\logg} and {\logteff}.  Two of these relations are included in Fig. \ref{fig:ttau}, which illustrates the effect of metallicity that is not captured in the Eddington or semi-empirical {\ttau} relations.  These two simulations ($s4$ and $s13$) have the same {\teff} and span the largest range in metallicity in the grid.

\section{Calibrated Stellar Models} \label{sec:method}

Stellar models with convective envelopes can be constructed if the chemical composition, mass, age, and mixing length parameter are specified.  Thus, for a star with a particular composition, the stellar model is characterized by three parameters.  It is often convenient to substitute the stellar surface parameters of {\logg} and {\logteff} for two of the three parameters in the (M,R,$\alpha$) triplet.

To determine how the {\ttau} relation affects the relationship between the mixing length parameter and metallicity, we compute sets of models corresponding to a particular set of atmospheric parameters and boundary conditions.  In effect, we create models for a star of a given {\logg} and {\logteff}.  Without additional constraints the model mass and radius are not unique and depend on the value of the mixing length parameter. 

Calibrated stellar models are computed in an iterative manner, using the Yale Stellar Evolution code  \citep{2008Ap&SS.316...31D} which uses the B\"ohm-Vitense formulation of the MLT.  For a given composition, stellar mass and surface gravity, the model is evolved until the desired radius (determined from stellar mass and {\logg}) is achieved.  The evolution is repeated with a different value for the mixing length parameter until {\logteff} also matches the desired value.  This processes is repeated for all of the stellar masses that yield solutions corresponding to the {\logg} and {\logteff} of the simulations in Table \ref{tab:sims}.   In order to keep the widest range of possible stellar masses, the models are not restricted in age.  This leads to some of the models being older than the age of the universe, and not physically realistic.  We include all of the models in our analysis because we aim to understand the behavior of the models, and we are not modeling an actual star.

The input physics in the stellar evolution code is consistent with that of the 3D simulations. We use the OPAL equation of state \citep{2002ApJ...576.1064R} and high-temperature opacities from \citet{1996ApJ...464..943I} along with low-temperature opacities from \citet{2005ApJ...623..585F}.  The models do not include convective core overshoot or the diffusion of heavy elements.
We compute calibrated stellar models with {\logg} and {\logteff} that correspond to the set of 3D radiation hydrodynamic (RHD) simulations.  In addition to the static {\ttau} relations that are typically used in stellar models, we can substitute the time-averaged atmospheric structures from the 3D simulations.

\begin{figure*}
\epsscale{1.15}
\plotone{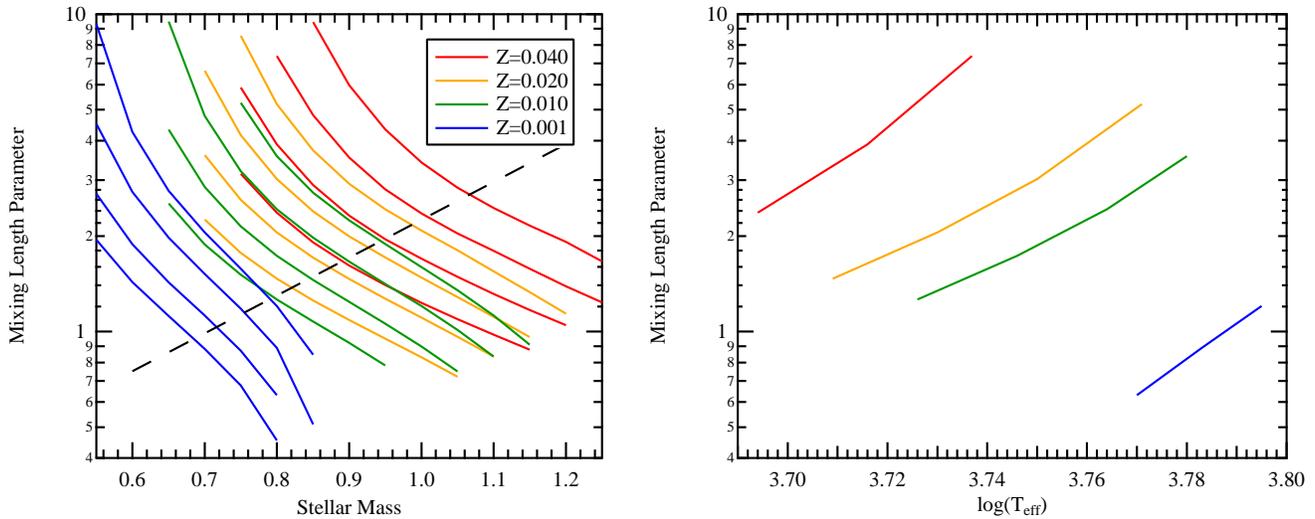}
\caption{\textit{Left:} \rev{Calibrated stellar models computed with the Eddington {\ttau} relation, corresponding to {\logg =4.30} and {\logteff} of the simulations in Table \ref{tab:sims}.  These models are not intended to represent a particular stellar population, and some of them are not physically realistic (models with ages less than that of the universe are below the dashed line).  
Comparing these models illustrates the behavior of MLT used in conjunction with a prescribed static {\ttau} relation.} \textit{Right:} A subset of the calibrated models with a stellar mass of $0.80 M_\odot$.} 
\label{fig:calibrated_models}
\end{figure*}

A range of stellar masses (and mixing length parameters) are possible for calibrated models with a given composition, {\logg} and {\logteff}.  In order to match the desired {\logg} and {\logteff}, smaller values for $\alpha$ are required at lower metallicity to compensate for the shifting of the evolution tracks to hotter effective temperatures.  The left side of Fig. \ref{fig:calibrated_models} shows the Eddington {\ttau} calibrated models corresponding to the simulations in Table \ref{tab:sims}.  Each curve corresponds to a particular effective temperature and metallicity.  Isolating a particular stellar mass and radius (right side of Fig. \ref{fig:calibrated_models}) reveals a more apparent metallicity-$\alpha$ trend for a fixed effective temperature.  For our analysis, we compute similar sets of calibrated models for different fixed {\ttau} relations, as well as with those from 3D simulations, which vary with $Z$ and {\logteff}.  The different sets of calibrated models reveal whether the metallicity-$\alpha$ trend depends on the choice of {\ttau} relation for the stellar models.

The thermal structure of the atmosphere in 3D simulations depends to some degree on the details of the radiative transfer solver.  For example, \citet{2012ApJ...759..120T} show that different radiative transfer schemes can yield differences of 20\% in the density through the superadiabatic layer.  Whether the differences are important will depend on how the 3D stratification is applied to stellar models.  To test if the radiative transfer solver in the 3D simulations affects the {\ttau} relations in a way that is significant, we duplicated two of the simulations using an alternate radiative transfer scheme.  The simulations listed in Table \ref{tab:sims} were all computed using the 3D Eddington approximation \citep{1966PASJ...18...85U}, but we re-computed simulations $s4$ and $s13$ with a long-characteristic ray integration method. 
We refer the reader to \citet{2012ApJ...759..120T} for a detailed description and comparison of these two radiative transfer schemes.  After switching the radiative transfer solver, simulations $s4$ and $s13$ were evolved for several thermal timescales to ensure that they were properly relaxed, at which time statistics were gathered in a manner identical to the other simulations.

\begin{figure*}
\epsscale{1.15}
\plotone{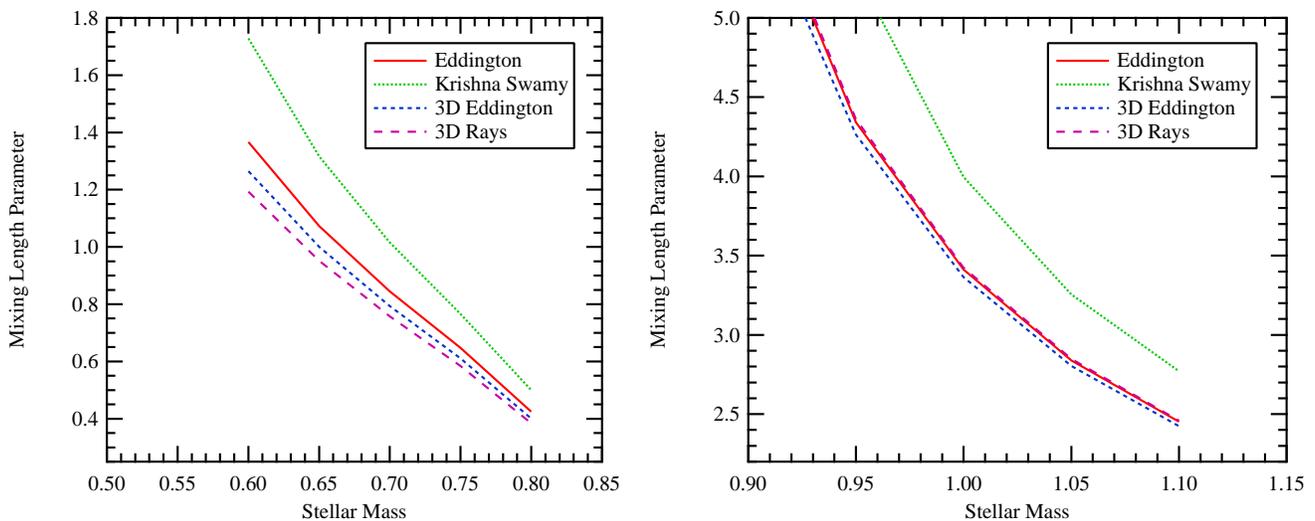}
\caption{\rev{A comparison of calibrated stellar models computed with different {\ttau} relations to show the relationship between the mixing length parameter and the surface boundary condition.}  All of the models have the same {\logg} and {\logteff} but differ in metallicity between the left ($Z=0.001$) and right ($Z=0.040$) panels. Two of the {\ttau} relations are extracted from 3D simulations that used alternative radiative transfer schemes. The relative effect introduced by changing the 3D radiative transfer scheme is more pronounced in the low metallicity case, but not large enough to significantly change the metallicity-mixing length relation.} 
\label{fig:rtmodels}
\end{figure*}

Stellar models computed with simulated {\ttau} relations show that the 3D radiative transfer scheme has a small effect on the mixing length determination.  Presented in Fig. \ref{fig:rtmodels}, the mixing length value as a function of stellar mass is shown for low ($Z=0.001$) and high ($Z=0.040$) The relative effect is somewhat larger at low metallicity, but is still too small to alter the relationship between the metallicity and the mixing length parameter.  The stellar mass ranges of the calibrated models differ between the two panels because the metallicities are quite different.

\section{Results and Discussion}\label{sec:trends}

To determine whether the stellar model surface boundary condition significantly affects the metallicity-mixing length trend reported by \citet{2012ApJ...755L..12B}, we compute several groups of calibrated models similar to those presented in Figure \ref{fig:calibrated_models} but with different surface boundary conditions.

\begin{figure}[h]
\epsscale{1.15}
\plotone{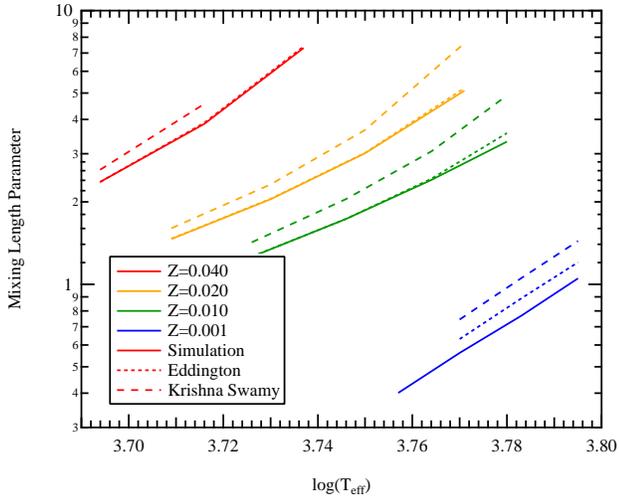}
\caption{Calibrated stellar models with a stellar mass of 0.80$M_\odot$ for several {\ttau} relations.  The metallicity-mixing length trend is clear and independent of the choice of {\ttau} relation.  \rev{The effect of introducing a metallicity-dependent {\ttau} relation is not large enough to change the overall $\alpha$-metallicity trend, although the relative effect on the mixing length parameter is larger at low-$Z$. }} 
\label{fig:allm08models}
\end{figure}

We make a $0.80 M_\odot$ cut in the set of models (presented in Fig. \ref{fig:allm08models}) to show the metallicity-mixing length trend.  The trend is clearly visible, with all low-Z models having smaller mixing length values than those with higher-Z.  Changing the {\ttau} relation from Eddington to KS introduces a shift in the mixing length value, but leaves the trend almost unchanged.  Introducing simulated {\ttau} relations shifts the mixing length values as a function of metallicity, and the effect is largest at low-Z.  Models that used simulated surface boundary conditions relations are quite similar to the Eddington models because both {\ttau} relations are similar at the stellar photosphere.

As mentioned in the introduction, the value of the mixing length parameter is sensitive to the atmospheric boundary condition of the stellar models.  To illustrate this effect we compare the set of Eddington {\ttau} calibrated models with those that have the KS {\ttau} relation.  The change in the surface boundary condition results in a shift in the mixing length parameters, shown with dotted lines in Fig. \ref{fig:allm08models}.     Switching the surface boundary condition between the Eddington and KS {\ttau} relations, however, does not change the basic behavior as a function of effective temperature or metallicity.

\citet{2012ApJ...755L..12B} performed a trilinear fit to their data to measure the variation of  the mixing length value as a function of {\logg}, {\logteff}, and [Fe/H].  Since our simulated data points share the same {\logg=4.30}, we perform a similar bilinear fit.
\begin{equation}
\alpha = a + b \log T_{\rm eff} + c[Fe/H],
\end{equation}
where we define the metallicity as:
\begin{equation}
[Fe/H] = \log(Z/X) - \log(Z/X)_\odot.
\end{equation}

Models were weighted for the linear regression such that the models comprising each metallicity group sum to equal values.  The result of the bilinear fit shows a robust correlation between metallicity and the mixing length value, regardless of whether a constant {\ttau} relation is used, or one that is a function of metallicity and derived from simulations.  The proportionality factor is $c=0.31 \pm 0.048$ for the complete sample, and $c=0.33 \pm 0.029$ for the age-restricted subset of models.

In the \citet{2012ApJ...755L..12B} study, the stellar masses and radii were asteroseismically determined.  In our set of models, it is possible that there is a $\alpha$-dependence on stellar mass, since our models span a range of stellar masses for a given {\logg} and {\logteff}.  In addition to the bilinear fit described above, we also performed a trilinear fit of:
\begin{equation}
\alpha = a + b \log T_{\rm eff} + c[Fe/H] + dM/M_\odot .
\end{equation}
The proportionality constant from this fit is $c= 0.74 \pm 0.019$ and $c= 0.64 \pm 0.030$ for the complete sample and  age-restricted subset, respectively.

The results of our bilinear and trilinear fitting are robust, and indicate that the relationship between metallicity and the mixing length value, as reported by \citet{2012ApJ...755L..12B}, are not an artifact of the stellar model surface boundary condition.  This strengthens the understanding that the solar-calibrated value for the mixing length parameter is not suitable for modeling other stars.  Although the metallicity-mixing length trend is significant, it is not possible to directly interpret the mixing length value, as it depends on the details of the formulation of MLT in the stellar model as well as on any input physics that affect entropy, of which the {\ttau} relation is just one.  For a given MLT formulation and set of input physics, changes in the mixing length parameter are still informative.  

We have carried out our tests on stellar models within the mixing length framework because MLT-like treatments of convection are predominant in stellar modeling today.   It is important to note, however, that the MLT formalism and the tunable parameter associated with it is just a proxy for describing convection in stars.  Even if a value for the mixing length can be extracted directly from 3D simulations \citep[e.g.][]{1999A&A...346..111L, 2011ApJ...731...78T}, MLT cannot correctly reproduce the properties of 3D simulations in the superadiabatic layer near the stellar surface.   

Going beyond the surface boundary condition to include other aspects of stellar convection may change the result as well.  This work focuses on the {\ttau} relation, but it contains only a small part of the information present in 3D simulations.  Including a representation of additional physical processes, such as the turbulent pressure contribution to hydrostatic equilibrium, will potentially yield different results.   It is desirable to ultimately replace MLT-like treatments of convection with a better description of stellar convection (for example, the ongoing efforts of \citealp{2013ApJ...769....1V}), and such models may behave quite differently than our current MLT models.  

\acknowledgments

JT is supported by NASA ATFP grant \#NNX09AJ53G to SB, and acknowledges a PGS-D scholarship from the Natural Sciences and Engineering Research Council of Canada. SB also acknowledges NSF grant \#AST-1105930. This work was supported in part by the facilities and staff of the Yale University Faculty of Arts and Sciences High Performance Computing Center.

\end{document}